\documentclass[12pt,letterpaper]{article}
\pdfoutput=1

\usepackage{graphicx,array}
\usepackage[dvipsnames]{xcolor}
\definecolor{darkblue}{rgb}{0,0.1,0.5}
\definecolor{darkgreen}{rgb}{0,0.5,0.2}
\definecolor{darkred}{RGB}{153,26,0}
\usepackage{ulem}
\definecolor{seablue}{rgb}{0,0.2,0.6}
\usepackage{latexsym}
\usepackage{amssymb, amsmath}
\usepackage{bm}        
\usepackage[numbers,sort&compress]{natbib}
\usepackage{bm,relsize}
\usepackage{slashed}

\definecolor{viola}{RGB}{134,41,198}
\usepackage{mathrsfs}
\usepackage{hyperref} 
\hypersetup{
    colorlinks=true,       
    linkcolor=darkblue,          
    citecolor=BrickRed,        
    filecolor=magenta,      
    urlcolor=MidnightBlue           
}
\usepackage[all]{hypcap} 
\usepackage{subcaption}

\usepackage[font=small,labelfont=bf,labelsep=period]{caption}

\newcommand{\GeV}{\mathrm{GeV}}

\newcommand{\Mpl}{M_{\rm Pl}}

\newcommand{\be}{\begin{equation}}
\newcommand{\ee}{\end{equation}}

 \date{\today}

\begin{document}

\begin{flushright}

\end{flushright}
\vspace{.6cm}
\begin{center}
{\LARGE \bf 
Gravitational Production of\\
\vspace{0.2cm}
a Conformal Dark Sector
}\\
\bigskip\vspace{1cm}
{
 Michele Redi$^{a,b}$, Andrea Tesi$^{a,b}$, Hannah Tillim$^c$}
\\[7mm]
 {\it \small
$^a$INFN Sezione di Firenze, Via G. Sansone 1, I-50019 Sesto Fiorentino, Italy\\
$^b$Department of Physics and Astronomy, University of Florence, Italy\\
$^c$Rudolf Peierls Centre for Theoretical Physics, University of Oxford,\\
Beecroft Building, Oxford OX1 3PU, United Kingdom
 }

\end{center}

\bigskip \bigskip \bigskip \bigskip

\centerline{\bf Abstract} 
\begin{quote}
Dark sectors with purely gravitational couplings to the Standard Model are unavoidably populated from the SM plasma by graviton exchange, and naturally provide dark matter candidates.
We examine the production in the relativistic regime where the dark sector is approximately scale invariant, providing general analytical formulas that depend solely on the central charge of the dark sector.  We then assess the relevance of interactions that can lead to a variety of phenomena including thermalisation, non-perturbative mass
gaps, out-of-equilibrium phase transitions and cannibalism in the dark sector.  As an illustrative example we consider the dark glueball scenario in this light and show it to be a viable dark matter candidate due to the suppression of gravitational production. We go on to extend these results to strongly coupled CFTs and their holographic duals at large-$N$ with the dark dilaton as the dark matter candidate.
\end{quote}

\vfill
\noindent\line(1,0){188}
{\scriptsize{ \\ E-mail:\texttt{ \href{mailto:michele.redi@fi.infn.it}{michele.redi@fi.infn.it}, \href{andrea.tesi@fi.infn.it}{andrea.tesi@fi.infn.it}, \href{hannah.tillim@physics.ox.ac.uk}{hannah.tillim@physics.ox.ac.uk}}}}
\newpage

\tableofcontents

\setcounter{footnote}{0}


\section{Introduction}

The entirety of the observational evidence for Dark Matter (DM) rests on its gravitational effects. While it is interesting to include non-gravitational interactions between DM and the Standard Model (SM), and certainly attractive to construct predictive models that can be tested directly, observational evidence increasingly indicates the existence of such couplings may be to a large extent wishful thinking. 
It is thus of utmost importance to establish and explore the most compelling scenarios of gravitationally coupled DM, in order to understand whether any testable prediction might be extracted.

At first sight obtaining predictions from a completely dark sector seems a daunting task, and one might have the impression that any model at all would reproduce observations.
In this note we will focus on the simplest scenarios where gravitational interactions are responsible for the population of the dark sector, assuming that the sector is not reheated by inflaton decay.  
Within these minimal assumptions the scenarios turn out to be rather predictive. We will consider  approximately conformal sectors such as gauge theories or strongly coupled CFTs, 
showing that the production  is determined only by the SM reheating temperature and the central charge of the dark sector. 
Gravitational production has been explored starting with \cite{Garny:2015sjg}, see also \cite{Tang:2016vch,Garny:2017kha,Tang:2017hvq,Gondolo:2020uqv,Chianese:2020yjo,Chianese:2020khl}, mostly focusing on free theories.  We will reproduce some known results, generalizing to interacting theories and showing the relevance of interactions for the abundance in these sectors.

The dark sector under discussion can only be produced from the SM plasma via graviton exchange\footnote{For a conformally coupled sector production through inflationary and post-inflationary fluctuations \cite{Chung:1998zb} is strongly suppressed. We also do not consider the possibility that the dark sector arises from evaporation of primordial black holes \cite{Lennon:2017tqq,Bernal:2020kse} as this requires additional inputs. That mechanism would provide a different initial condition for conformal sectors that could then be studied along the lines this work. Several works also studied CFTs connected to the SM through other mediators or effective operators see for example \cite{Harling:2008px,vonHarling:2012sz,Katz:2015zba,Hong:2019nwd}.}. 
Let us note that even though the process happens at tree level,  gravitational production relies on the quantum nature of gravity.
The dark sector is populated by the freeze-in mechanism \cite{Hall:2009bx} from the hot SM plasma, which itself is well approximated by a thermal CFT. Due to the non-renormalizable nature of gravitational interactions, particle production is most effective at high temperatures. If the reheating temperature of the visible sector is sufficiently large, this dark sector can provide the entirety of the DM abundance. In recent literature there has often been a focus on dark sectors produced gravitationally but without additional self-interactions (see however \cite{Garny:2018grs,March-Russell:2020nun,Bernal:2020ili,Bernal:2020kse}). In this work, we discuss in contrast interacting approximately conformal field theories (CFTs) coupled through gravity showing that, in many cases, self-interactions allow the dark sector to reach thermal equilibrium with a temperature much lower from that of the visible sector.

We focus on two simple examples. In the first we consider a pure gauge Yang-Mills theory decoupled from the SM. Before confinement, the theory is approximately scale-invariant and enjoys Weyl symmetry. Upon confinement at a first order phase transition, the theory develops a mass gap and the lightest states are spin-0 dark glueballs. We discuss the gravitational production in the deconfined regime, then explore the dark QCD sector's evolution towards thermal equilibrium, in particular how the latter is affected by the phase transition. We find that, thanks to the gravitational production, it is possible to realize a viable scenario of glueball DM, overcoming the overproduction of  DM that typically hinders this scenario.

Next we consider the case of a strongly coupled CFT, where the DM is the dilaton associated to the spontaneous breaking of conformal invariance in the dark sector. In this context, we find that the confinement/deconfinement phase transition differs from the standard picture in light of the different temperatures of the visible and dark sectors - this has a calculable impact on the DM abundance. We also find in this case  that the dark dilaton can be a good DM candidate.

We wish to highlight the main novelties of this work. In section \ref{sec:production} we derive in full generality the abundance of a gravitationally coupled dark sector using conformal field theory techniques and unitarity. This elegantly reproduces known results for weakly coupled theories allowing to also study strongly coupled sectors where conventional methods are not applicable. In section \ref{sec:glueballs} we study for the first time the gravitational production of pure glue dark sector where the lightest glueball is DM. When the gluons thermalize this connects to previous works on glueball DM \cite{Cline:2013zca,Boddy:2014yra,Soni:2016gzf,Forestell:2017wov,Acharya:2017szw,Jo:2020ggs}, explaining why the dark sector temperature is lower than the visible one as demanded on phenomenological grounds. We also study the novel possibility that the dark sector does not thermalize leading to out of equilibrium phase transitions. In section \ref{sec:CFT} we study the analogous problems at strong coupling focusing on the scenario where DM is a dark dilaton coupled to the SM only through gravitational interactions.
In both cases we discuss the phases of the theory and the DM abundance and constraints. 
We summarize our results and outline future directions in section \ref{sec:conclusions}. In the appendix \ref{app:A} we provide a general derivation for the gravitational production of CFTs and provide the solution of the Boltzmann equation that determines the phase space distribution at production.

\section{Gravitational production of Dark Sectors}\label{sec:production}
The universality of gravitational interactions provides an unavoidable source of DM. Two mechanisms have been proposed in the literature:
production from the SM plasma through tree-level graviton exchange \cite{Garny:2015sjg} and production through quantum fluctuations in an expanding background \cite{Chung:1998zb}. 
Here we will focus on the first production mechanism as the latter requires explicit breaking of conformal invariance that is very suppressed in our models. Indeed in the relativistic regime 
breaking of conformal invariance is in fact non generic. In particular, massless gauge theories and fermions are automatically conformally invariant. 
Moreover tree-level gravitational production is very predictive being determined by the reheating temperature.

\subsection{Tree level production}

Gravitational interactions are too weak to establish thermal equilibrium between the dark sector and the SM.
Nevertheless, particles in a dark sector are inevitably produced through gravitational interactions at tree level, thanks to the exchange of a graviton in the $s$-channel. We assume for simplicity that the visible sector is reheated (instantaneously) to a temperature $T_R$, and the Hubble rate is dominated by contributions from the SM thermal bath\footnote{The assumption of instantaneous reheating  allows us also to relate initial conditions in terms of the Hubble scale during inflation, $H_I$, being $3 H_I^2 \Mpl^2=\pi^2 g_* T_R^4/30$. Using the latest results from PLANCK and BICEP2, the bound on the scalar to tensor ratio $r<0.07$ implies $H_I<7 \times 10^{13}$ GeV. This translates into an upper bound $T_R\lesssim 0.7\times 10^{16}\, \GeV $.}.

Pairs of SM particles can annihilate into dark sector states via the exchange of a graviton, and the amplitude for this transition reads in general
\be
\mathcal{A}=\frac{1}{\Mpl^2 s} \bigg( T_{\mu\nu}^{\rm SM}T_{\alpha\beta}^{\rm DM}\eta^{\mu\alpha}\eta^{\nu\beta} - \frac12 T^{\rm SM} T^{\rm DM}\bigg)\,.
\ee

In the case of a dark conformal sector, with $T_{\rm DM}=0$, only the first term inside the brackets contributes.
The rate is suppressed by the gravitational coupling, here $\Mpl=2.4\times 10^{18}\GeV$, so that in order to populate the dark sector one needs to rely on very high reheating temperatures.  
The Boltzmann equation for the yield $Y_D=n_D/s$ is given by\footnote{We neglect quantum statistics for the dark sector, allowing us to derive analytical formulas for the energy distribution. The relevant dynamics takes place at  temperatures larger than 100 GeV so that we take $g_*=106.75$ throughout.}
\begin{equation}
 \frac{dY_D}{dT} = \frac{\langle \sigma v\rangle s(T)}{H T}(Y_D^2- Y_{\rm eq}^2)\,,~~~~~~~~~~~~~~~~~~~~~Y_{\rm eq} =\frac {g_{\rm D}}{g_*} \frac{45}{2\pi^4}
\label{eq:boltzmann}
\end{equation}
where  $\langle \sigma v\rangle$ is the thermally averaged annihilation cross-section of DM summed over SM final states:
\be
\langle \sigma v\rangle= 4 \langle \sigma_0 v\rangle +45  \langle \sigma_{1/2} v\rangle + 12\langle \sigma_1 v\rangle.
\ee

Assuming a negligible initial density the Boltzmann equation has the following approximate solution at late times:
\be
Y_D(0)= \int_0^{T_R} \frac{dT}{T} \, \frac{\langle \sigma v\rangle s}{H}Y_{\rm eq}^2\,.
\ee
valid as long $Y_D\ll Y_{eq}$, which as we will show is always be satisfied.
In the above formula there are a couple of assumptions. As previously stated, we consider an instantaneous reheating that allows us to neglect the details of the inflaton sector above $T_R$. Second, since the interactions are non-renormalizable, we integrate down to $T=0$ without introducing errors given that the production is strongly dominated by the highest temperature.

The exact expression for $\langle \sigma v\rangle$ will depend on the details of the dark sector theory considered. 
However, for a conformally-invariant dark sector and for $T_R$ much above the electroweak scale, the cross-section can be written as
\be
\langle \sigma_i v\rangle = \frac {c_i c_{D}}{g_D^2} \frac {3}{1280\pi} \frac{T^2}{\Mpl^4}\,.
\ee
Here $c_D$ and $c_i$ are the central charges of the dark sector and of the SM particle and $g_D$ the number of degrees of freedom of the species. 
This formula is completely general and can be applied to the gravitational production of any 
CFT state. \footnote{Let us note that this formula even applies to production of gravitons. 
If the initial states are conformally coupled the yield is  given by eq. (\ref{eq:yield}) with an effective central charge $c_2=28$. The only exception to this formula comes from the special case where initial and final states are not conformally coupled. For scalars with couplings to curvature $\xi/2 \phi^2 R$, the cross-section receives extra contribution $3/(4\pi)(\xi_i-1/6)^2 (\xi_D-1/6)^2 T^2/\Mpl^4$.}
Integrating (\ref{eq:boltzmann}) we find
\begin{equation}
Y_{\rm D} = a_{\rm SM} \, c_{D} \bigg(\frac{T_R}{\Mpl}\bigg)^3\,,~~~~~~~~~~~~~~~~a_{\rm SM}= \frac{27 \sqrt{5/2}}{256 \pi^8 g_*^{3/2}}\times \sum_i c_i\,.
\label{eq:yield}
\end{equation}
For a real scalar, Weyl fermion and massless gauge field the values of the central charges are \cite{Osborn:1993cr}
\begin{equation}
c_0=  \frac 4 {3} \,,~~~~~~~~~~~~~~~~~~~c_{1/2} = 4\,,~~~~~~~~ c_1= 16\,.
\end{equation}
Numerically one finds $a_{\rm SM}=6\times 10^{-6}$.
From these results we can derive the deviation from thermal equilibrium in the dark sector,
\be
\frac{n_{\rm D}}{n_{\rm eq}}\approx 0.0028 \frac {c_D}{g_D} \left(\frac{T_R}{\Mpl}\right)^3\,.
\ee

Gravitational production gives rise to a dark sector underpopulated compared to thermal equilibrium.
The typical energy of the dark quanta produced are however of order of the temperature of the visible sector so that the energy density is of order
$\rho_D\approx c_D T^4 T_R^3/\Mpl^3$. We can also perform a more refined computation by solving the Boltzmann equation for the phase distribution of DM particles. 
This computation can be found in appendix \ref{app:A}. Neglecting small corrections from quantum statistics, one finds
\begin{equation}
f_D(T,p)\approx \frac  {2\pi^4 g_*} {135} Y_D \frac {p \,  e^{-p/T}}T\,.
\label{eq:DMdistribution}
\end{equation}

As expected, since the cross-section grows with energy, the distribution slightly favours higher energies compared to the Boltzmann distribution. From eq. (\ref{eq:DMdistribution}) we can derive the energy density which can be cast in the following form
\be
\rho_D= \kappa\, \frac{\pi^2}{30} g_{\rm D} T^4 \, \frac{n_{\rm D}}{n_{\rm eq}}\,,
\ee
where the coefficient $\kappa=120/\pi^4 \approx 1.23$. 

It is useful to introduce the ratio of the energy densities of the visible and dark sectors,
\be\label{eq:ratiodensity}
r \equiv \frac{\rho_D}{\rho_{\rm SM}} =\kappa\,\frac{g_{\rm D}}{g_*}\, \frac{n_{\rm D}}{n_{\rm eq}}\approx 0.0035\frac{c_D}{g_*} \left(\frac {T_R}{\Mpl}\right)^3\,.
\ee

This quantity is conserved as long as the system is relativistic and decreases as $1/a$ when the dark sector becomes non-relativistic.
It has a similar physical meaning to the ratio of entropies of dark and visible sectors that however strictly cannot be defined out of equilibrium.

\subsection{Dynamics after production}
\label{sec:dynamics}

What happens next depends on the interactions in the dark sector.

\paragraph{No Interactions:}
This is the scenario studied in \cite{Garny:2015sjg,Garny:2017kha}. 
If the interactions are sufficiently weak, i.e. the rate of the relevant processes $\Gamma\sim n_D \sigma v\ll H$,
$Y_D$ remains approximately constant after production, we use eq. (\ref{eq:yield})
to obtain the DM abundance
\be\label{eq:DMnoint}
\frac{\Omega h^2 }{0.12}=  \frac{Y_D M}{0.4\, {\rm eV}}\approx  \frac {c_D M}{10^6\,{\rm GeV}} \left(\frac{T_R} {10^{15}\,{\rm GeV}}\right)^3\,,
\ee
where in this context the DM mass $M$ arises via a spontaneous breaking of scale invariance or from other mechanisms such as confinement. 
The main assumption is that it plays no role in the gravitational production at temperature $T_R\gg M$. 

\paragraph{Kinetic  and chemical equilibrium:}
We introduce interactions in the dark sector. Naively at weak coupling the first interactions that become relevant are $2\to 2$ processes. 
These transitions do not change the number density in the dark sector but allow the system to approach kinetic equilibrium. If kinetic equilibrium is reached, 
the phase distribution develops a chemical potential that can be determined using number density and energy density conservation.
Neglecting for simplicity quantum statistics one finds
\be
f=  e^{-\frac{E-\mu}{T_D}} \longrightarrow T_D= \frac 4 3 T\,, ~~~e^{\frac {\mu}T}\approx  \frac{27}{64} \frac {n_D}{n_{\rm eq}}\,.
\ee

When number-changing processes such us $2\to3$ become relevant the system can reach full chemical equilibrium. 
It is found in fact that due to collinear singularities such processes might be even faster \cite{Garny:2018grs} than elastic ones, leading
to faster thermalization. For simplicity we parametrize the cross-section in the massless limit as $\sigma_{2\to 3}= \alpha_{\rm eff}^3/T^2$ so 
the rate $\Gamma= n_D \sigma_{2\to 3}$ exceeds the Hubble rate for visible sector temperatures
\begin{equation}\label{condition}
T^* \approx 8 \cdot 10^{-5} (c_D \alpha_{\rm eff}^3) \left(\frac{T_R}{\Mpl}\right)^3 \Mpl \approx 1.4 \times 10^4\,{\rm GeV} (c_D \alpha_{\rm eff}^3)\left(\frac{T_R}{10^{15}\,{\rm GeV}}\right)^3
\end{equation}

When this happens we can directly compute the dark sector temperature from  the conservation of the energy density in the dark sector, eq.  (\ref{eq:ratiodensity}):
\be\label{ratioT}
\frac{T_D}{T}=\bigg(\frac{g_{*}r}{g_{\rm D}}\bigg)^{\frac14}\, \approx 0.25 \left(\frac {c_D}{g_D}\right)^{\frac14} \bigg(\frac{T_R}{\Mpl}\bigg)^{\frac34}\,,
\ee
so that the dark sector thermalises  at temperature
\begin{equation}
T_{D}^*\approx  \,  10\, {\rm GeV}\, c_D \alpha_{\rm eff}^3  \left(\frac {c_D}{g_D}\right)^{\frac14}  \left(\frac{T_R}{10^{15}\,{\rm GeV}}\right)^{\frac {15}4}
\label{eq:TD*}\, .
\end{equation}
Consistency requires that the mass scale is smaller than $T_D^*$ to justify the relativistic computation. 
After thermalisation the yield becomes
\begin{equation}
Y_{\rm th}= \frac{2 \zeta(3) g_D}{45\pi^4 g_*}  \left(\frac {T_D}T \right)^3 \approx 8 \cdot 10^{-8} g_D \left(\frac {c_D}{g_D}\right)^{\frac34} \left(\frac{T_R}{\Mpl}\right)^{\frac {9}4}\,.
\end{equation}

Using eq. (\ref{eq:yield}) we can see that the yield is increased by a factor $10^{-2} (g_D/c_D)^{\frac14} (M_p/T_R)^{\frac34}$ compared to the free case.

When the dark sector thermalises it acquires a much lower temperature than the SM and starts  its own thermal history. 
Observable quantities are thus rescaled by the ratio $T_D/T$.

\paragraph{Out-of-equilibrium dynamics:}
If the interactions are not sufficiently strong, $T_D^*< M$ and the system does not reach thermal equilibrium in the relativistic regime.
In this case the interactions can still modify the abundance. Moreover, if the system undergoes a phase transition such as confinement,
very unconventional dynamics are realized. In our case we expect the phase transition to take place when the correlation length becomes comparable to the dynamical scale, $n_D\sim \Lambda^3$.
At this point the quanta in the deconfined phase have an energy $E\gg \Lambda$. The situation looks similar to that occurring in collider physics, where energetic gluons and quarks move in the
confined vacuum and hadronize. This significantly modifies the abundance since each perturbative state generates a large number of DM particles.
In light of the enhanced number density, interactions might still be able to thermalise the sector in the confined phase, giving rise to cannibalistic effects \cite{Carlson:1992fn,Pappadopulo:2016pkp,Farina:2016llk}.

In what follows we present explicit examples of the above in strongly gauge-coupled conformal sectors based on the AdS/CFT correspondence.
 
\section{Dark Glueballs}
\label{sec:glueballs}
The simplest dark sector conformally coupled to gravity is a theory of pure Yang-Mills in the perturbative regime.
This sector confines at a scale $\Lambda_{\rm DC}$, forming color singlet glueballs. The lightest glueball is stable except for decays to gravitons and  thus provides 
an attractive DM candidate  \cite{Cline:2013zca,Boddy:2014yra,Soni:2016gzf,Forestell:2017wov,Acharya:2017szw,Jo:2020ggs}. This scenario however requires the temperature of the dark sector, if ever in thermal equilibrium, 
to be much smaller than that of the SM. This condition might appear unnatural but in fact we will show that it is automatically realized through gravitational production.
Moreover the non-perturbative interactions of gluons and glueballs are crucial to determine the DM abundance. 
 
For concreteness, we consider a pure glue theory with gauge symmetry SU($N$).
If $T_R\gg \Lambda$, the sector is relativistic at production and consists of a theory of perturbative gluons.
The gravitational production from the SM plasma can be computed using eq.~\eqref{eq:yield}, where
\be
c_{\rm D}=16(N^2-1) \,.
\ee

After production in the perturbative regime the gluon energy and density redshifts due to the expansion and interactions unavoidably become relevant, leading eventually 
to color confinement. If the system thermalises in the relativistic regime the phase transition is the standard thermal phase transition of strongly couple gauge theories but it is also possible that the phase transition 
takes place out of equilibrium. In this case the transition occurs when the system has a number density smaller than $\Lambda^3$, i.e. the particles are separated by a distance larger than $1/\Lambda$. 
After the phase transition physical states are glueballs. Glueball interactions, in particular number changing processes, are again important in the region where the DM abundance is reproduced.

\begin{figure}[t]
\centering
\includegraphics[width=0.7\linewidth]{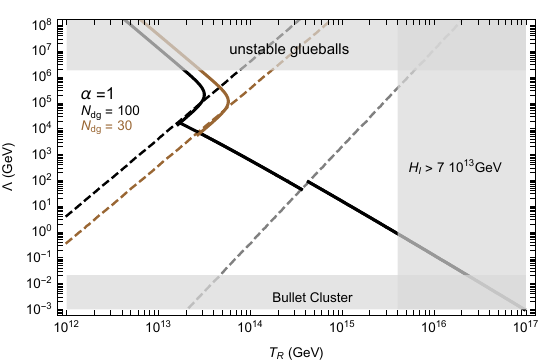}
\caption{\label{fig:cannibalism}\it Phase diagram of gravitationally produced glueball DM. DM abundance is reproduced on the black and brown lines. In the lower right region below the dashed grey line gluons thermalise in the relativistic regime and then undergo cannibalism, reducing their numerical abundance. In the middle region between dashed lines gluons confine before thermalisation but glueballs then thermalise. In the upper left region above the dashed lines the system never thermalises and the abundance depends on $N_{\rm DG}$ (black and brown lines). The grey regions are excluded by DM self-interaction, scale of inflation and lifetime of glueballs where we use the estimate $\tau_{\rm DG}\sim \Mpl^4/M_{\rm DG}^5$.}
\end{figure}

We now discuss in detail the scenario where the confinement phase transition takes place in or out of thermal equilibrium.
We will focus for simplicity on SU($N$) gauge theories with 3 colors ($c_D=128)$ where a wealth of lattice results are available.
In figure \ref{fig:cannibalism} we show a cartoon of the phase diagram of theory in the plane  and the values of $(\Lambda, T_R)$
where the DM abundance  is reproduced.

\subsection{Thermal Gluons}
The thermalisation of pertubative gluons has been studied in detail for applications to the  quark gluon plasma in the SM.
The main difference in our case is that there are no quark degrees of freedom and that the plasma is initially very diluted. 
Due to the long-range nature of the interactions the analysis in section \ref{sec:dynamics} should be refined and, 
strictly speaking, simply solving Boltzmann equations is not sufficient, see \cite{Arnold:2007pg} for an elementary introduction and \cite{Garny:2018grs} for a discussion in the context of
abelian gauge theories.  In cite \cite{Arnold:2002zm} the thermalization of a non-abelian gluon gas in the perturbative regime was studied using kinetic theory. In \cite{Kurkela:2014tea}
the thermalization of a diluted perturbative gas was also studied using numerical methods. In our scenario the details of thermalization are
not important as long as the process completes in the relativistic regime. The boundary of thermalization is determined by the non-perturbative region where the results above are not
directly applicable. Because of this we simply estimated the critical value of the temperature by taking $\alpha_{\rm eff}=1$.

For what concerns the critical temperature, given that it is determined by conservation of energy the details of transient are not important 
so that the estimate in eq. (\ref{ratioT}) is reliable. Using the simple parametrization in eq. (\ref{condition}), for $N=3$ we find the thermalisation temperatures
\begin{equation}
T^* \approx  10^6\,{\rm GeV} \alpha_{\rm eff}^3 \left(\frac{T_R}{10^{15}\,{\rm GeV}}\right)^3\, ~~~~~T_D^* \approx 10^3\,{\rm GeV} \alpha_{\rm eff}^3 \left(\frac{T_R}{10^{15}\,{\rm GeV}}\right)^{15/4}\, .
\label{eq:thermalgluons}
\end{equation}
Thermalisation in the relativistic regime is realized if $\Lambda< T_D^*$ where we can take $\alpha_{\rm eff}\sim1$ as the maximum value.

\paragraph{Confinement Phase Transition:}~\\
When the system reaches thermal equilibrium before the phase transition, the results will be the same as those of glueball DM, with the striking difference that the glueballs are now much colder than the SM sector. In order to determine the abundance of glueballs we can use lattice results for $SU(3)$ pure glue gauge theory (similar results apply for small $N$).
The mass of the lightest glueball and the latent heat of the phase transition are roughly \cite{Beinlich:1996xg,Morningstar:1999rf,Borsanyi:2012ve}
\begin{equation}
\frac {M_{\rm DG}}{\Lambda}\approx 5.5\,,~~~~~~~~~~~~~~\frac{L_h}{\Lambda^4}\approx 1.4\,,
\end{equation}
where $\Lambda$ is the critical temperature of the dark sector. 
Since the glueballs are rather heavy compared to the critical temperature they will be produced non-relativistically. 
Two possibilities arise in this case, depending on whether self-interactions of glueballs allow number-changing processes with a fast rate.

\paragraph{Free glueballs:}~\\
If number-changing processes were not efficient, the number density of glueballs could be simply calculated from the energy released during the phase transition. 
For confining gauge theories the latter is expected to be rapid so the nucleation temperature $T_n$ will be similar the critical temperature $\Lambda$ \cite{Garcia:2015loa}.
Assuming that the transition completes instantaneously from energy conservation we obtain, 
\be\label{DG-no-interactions}
Y_{\rm DG}=\frac{\rho_{th}(T) +L_h}{M_{\rm DG}\, s(T)}\bigg|_{T_n} \approx 0.01 \frac{\Lambda}{M_{\rm DG}} \left(\frac{T_R}{\Mpl}\right)^{9/4}\,,
\ee
where we estimated the energy density of gluon plasma in terms of free gluons and  used (\ref{ratioT}). From this we derive the DM abundance:
\begin{equation}
\frac{\Omega  h^2}{0.12}\approx \frac {M_{\rm DG} }{10\,{\rm GeV}} \left(\frac{T_R} {10^{15}\,{\rm GeV}}\right)^{9/4}\,.
\end{equation}

Note that DM relic abundance is achieved for much lower values of $M$ compared to the non-interacting case.
In practise however we find that in the region where DM abundance is reproduced number changing processes are relevant.

\paragraph{Cannibalism:}~\\
If number-changing processes are important, the yield can be modified from eq.~\eqref{DG-no-interactions}. In a theory of pure glueballs, the leading processes that change the number density are  $3\to 2$. These reactions have a cross section $\langle \sigma_{3\to 2} v\rangle\equiv \alpha_{\rm DG}^3/M_{\rm DG}^5$, and just after the phase transition they are faster than Hubble if $\alpha_{\rm DG}^3 (\Mpl/M_{\rm DG}) \sqrt{r}\gg 1$. If this condition is met, the glueballs maintain thermal equilibrium for a while, despite being non-relativistic, see for example \cite{Forestell:2017wov}.

These interactions reduce the number density of glueballs and increase their temperature, which only marginally drops due to the expansion. This behaviour is due to the fact that we are considering a non-relativistic plasma in thermal equilibrium at fixed entropy. The mechanism is often dubbed `cannibalism' \cite{Carlson:1992fn}, since DM `eats' itself in order to warm up. 
To estimate the effect of cannibalism  we follow the standard approach to consider $3\to 2$ interactions,  see \cite{Carlson:1992fn,Pappadopulo:2016pkp,Farina:2016llk,Cline:2013zca,Boddy:2014yra,Soni:2016gzf,Forestell:2017wov,Acharya:2017szw,Jo:2020ggs}. The Boltzmann equation for the abundance reads
\be
\frac{T}{Y_{\rm eq}}\frac{d Y}{dT} = \frac{\langle \sigma_{3\to 2} v\rangle n_{\rm eq}^2}{H} \bigg[\frac{Y^3}{Y_{\rm eq}^3}-\frac{Y^2}{Y_{\rm eq}^2}\bigg]\,,
\ee
where $n_{\rm eq}$ is the (non-relativistic) equilibrium distribution at a temperature $T_D$. This equation has to be complemented with one governing the behaviour of $T_D(T)$. At the phase transition the temperature of the new confined phased is determined by conservation of energy (assuming instantaneous thermalisation). This implies that the ratio of entropies of dark and visible sector
is,
\be
\rho(T_D^*)=\rho_{th}(T_n)+L_h\quad \longrightarrow\quad q\equiv \frac {s_d}{s}=\frac{\rho(T_D^*)}{T_D^*\,s}\,.
\ee
After the completion of the phase transition, the dark sector entropy is conserved. This allows to determine the evolution of the temperature of glueballs  \cite{Carlson:1992fn} 
\be\label{temp0}
q=\frac{M_{\rm DG}}{T_D}\frac{n_{\rm eq}(T_D)}{\frac{2\pi^2 g_*}{45}T^3}\,,\quad \to \quad \frac{T_D}{M_{\rm DG}}= \frac{2}{W\big[(T_0/T)^6\big]}\, \quad T_0\equiv 0.59 \frac{M_{\rm DG}}{g_*^{1/3} q^{1/3}}\,.
\ee
where $W(z)$ is the solution of $W e^W=z$. At small temperatures the approximate solution is simply $T_D/M_{\rm DG} \approx 1/(3 \log(T_0/T))$. The logarithmic redshift of the dark sector temperature is related to the fact that the system maintains thermal equilibrium despite being non-relativistic. In fact the equilibrium yield is now given $Y_{\rm eq}=q\, T_D/M_{\rm DG}$, which does not drop exponentially.

The freeze-out of cannibalism occurs when $\Gamma\approx \sigma_{3\to 2}n_{\rm eq}^2$ drops below Hubble at the temperature (inserting eq.~\eqref{temp0})
\be
T_f =2 a^{-1/4}\sqrt{W(a^{3/8}T_0^{3/2}/4)}\,, \quad a\equiv 2.32\frac{\alpha^3 g_*^{3/2}}{M_{\rm DG}^4}\frac{\Mpl}{M_{\rm DG}}\, q^2\,.
\ee

Approximately,
\be
\frac{T_D}{M_{\rm DG}}\bigg|_{T_f} \approx \frac{1}{3\log Q}\,, \quad Q\equiv 0.08 \displaystyle \left(\frac{g_*}{106.75}\right)^{1/24} \left(\frac{\alpha}{0.1}\right)^{3/4}\, q^{1/6} \left(\frac{\Mpl}{M_{\rm DG}}\right)^{1/4}\,.
\ee

The resulting yield after cannibalism is then given by
\be
Y_{\rm cannibal}\approx  \frac{1}{3\log Q}\,\frac{g_D}{g_*}\bigg(\frac{3}{4} + \frac{1.4\times 45}{2\pi^2 g_D}\bigg)\frac{\Lambda}{T_D^*}\bigg|_{T_n}\, \bigg(\frac{g_{*}}{g_{\rm D}}\bigg)^{\frac34}\, r^{\frac34}\,,
\ee
that is directly proportional to the ratio of entropy densities \cite{Carlson:1992fn}.

This is the typical behaviour of the energy density of a non-relativistic plasma in equilibrium with conserved entropy, $\rho_{\rm DG}\sim (a^3 \log a)^{-1}$. The cannibalism therefore prevents a larger redshift of the energy density and leaves the yield in eq.~\eqref{DG-no-interactions} essentially unaffected save for a small logarithmic correction.  We note that the mass scale $M_{\rm DG}$ does not play a large role in the determination of $Y_{\rm cannibal}$.

As shown in figure \ref{fig:cannibalism} due to cannibalism the DM abundance can be reproduced even for masses below GeV. 
In this regime self-interactions of glueballs become relevant. We estimate the elastic cross-section among glueballs to be geometric,
$\sigma_{el} =\pi /M_{\rm DG}^2$. The bullet cluster bound on self interactions $\sigma_{\rm el}/M_{\rm DG} < 2 \cdot 10^{3}\, {\rm GeV}^{-3}$ implies the estimate $M_{\rm DG}>0.1$ GeV.
Lattice results in SU(2)  give $\Lambda > 50$ MeV \cite{Yamanaka:2019yek}.

\subsection{Out-of-Equilibrium Gluons}
If the confinement scale $\Lambda > T_D^*$ in eq.~\eqref{eq:TD*}, the system undergoes confinement before thermalisation.
In such a case we expect the transition to take place when the typical distance between quarks is of order $\Lambda$\footnote{We assume here that $H< \Lambda$ so that the curvature of space does not play role. In the opposite regime the fast expansion of the universe forbids the transition.} which occurs at the visible sector temperature $T_\Lambda$ determined by $n_D(T)=Y_{\rm D} s(T)\sim \Lambda^3$
\be
T_\Lambda\approx \Lambda\,\bigg(\frac{\pi^2}{g_{\rm D} r }\bigg)^{\frac13}  \sim \Lambda \frac{\Mpl}{T_R}\,.
\label{eq:TLambda}
\ee

Note that, since the system failed thermalise,  $T_\Lambda$ is also the typical energy of gluons at the onset of the phase transition - 
the gluons have typical energy larger than thermal distribution by a factor $M_p/T_R$. 
This leads cosmological collider physics: each high energy gluon moving in the confined vacuum will hadronize, producing jets of color-singlet glueballs.
If each gluon produces on average $N_{\rm DG}$ dark glueballs the  yield of glueballs after the phase transition is
\be\label{glueballs-nonthermal}
Y_{\rm DG}(T_\Lambda) = N_{\rm DG}Y_{\rm D} \,,~~~~~~~~~~Y_D=0.0008 \left(\frac{T_R}{\Mpl}\right)^3\,.
\ee

The DM abundance is thus increased by $N_{\rm DG}$ compared to (\ref{eq:DMnoint}).  The number of glueballs is difficult to estimate
but will scale with $T_\Lambda/\Lambda$ which in light of eq. (\ref{eq:TLambda})  is expected to be constant for a given reheating temperature. 
For example, at the LHC a 1 TeV jet can contain $\mathcal{O}(100)$ hadrons. 
Note that in this case the energy released in the phase transition is negligible compared to the energy of the gluons.

However this is not the end of the story, since the glueballs undergo number changing processes and can themselves thermalise.
In this scenario, the glueballs are relativistic right after the phase transition, with a typical energy of the type $E\sim T/N_{\rm DG}$, so that the cross-section
of number changing processes scales as  $\sigma \sim N_{\rm DG}^2 \alpha_{\rm DG}^3/T^2$ if they are relativistic. Because the temperature dependence is identical to that in section
(\ref{sec:dynamics}) we can simply rescale the formulae to take into account $N_{\rm DG}$. Thermalisation of dark glueballs takes place 
at visible and dark sector temperatures given by eq. (\ref{eq:thermalgluons}) multiplied by $N_{\rm DG}^3$.
This shows that even if the gluons do not thermalise before confinement, 
the glueballs  may do so afterwards provided they are relativistic $T/N_{\rm DG}>\Lambda$. This translates to an upper bound on the confinement scale:
\be
\Lambda \lesssim  10^{6}\,{\rm GeV} \,\alpha_{\rm DG}^3 N_{\rm DG}^2 \left(\frac {T_R}{10^{15}\,{\rm GeV}}\right)^3\,.
\ee

When $\Lambda$ satisfies this constraint the glueballs thermalize while relativistic. This, as in the case of gluons discussed in section 3.1, is accompanied by an increase in number density 
and a suppression of the dark sector temperature. The  ratio of dark sector and visible temperature is again given by eq.~\eqref{ratioT}, with $g_D$ here replaced by the multiplicity of lightest glueballs. Once the system thermalizes the computation of the relic density follow the one of the previous section. In particular if number changing processes remain in equilibrium when the glueballs become non-relativistic a phase of cannibalism is realized also in this case. If instead the glueballs do not thermalise their abundance is roughly given by (\ref{glueballs-nonthermal}).

The different regions of gravitationally produced glueball DM are shown in figure \ref{fig:cannibalism}.

\section{Dark CFT}\label{sec:CFT}

In this section we investigate the possibility that the dark sector is a strongly-coupled conformal field theory with a relevant or  irrelevant deformation. 
We assume that the latter creates a mass gap $\Lambda$ in the infrared and that the lightest state, accidentally stable, is the DM.

The formulae in section \ref{sec:production} can be directly applied to this scenario if $T_R\gg \Lambda$ as they rely only on the central charge of the CFT, see Appendix \ref{app:A}. 
In general $g_D \sim c_D$ but it is not obvious how to precisely define the number of degrees of freedom of the CFT. Fortunately however $g_D$ drops out of the relevant formulae for energy and number density that are needed to determine the DM abundance. In contrast to weakly coupled theories, for general CFTs what is being produced are not conventional particles but `CFT shell states' of a given energy. 
Because the phase space distribution is similar to the one of a fractional number of particles this is also known as `unparticle production' \cite{Georgi:2007ek} (see \cite{Kikuchi:2007az} for an application to DM).

Let us also mention that for elementary scalars an ambiguity exists in the coupling  to gravity. This is due to the fact that $\phi^2 R$ is a dimension 4 operator
so that it can be added to the action to leading order. In interacting CFTs this is not expected to happen as, at least in supersymmetric cases, no scalar operator of dimension 2 exists.
This implies that the coupling to gravity is uniquely determined. A related fact is that the dark sector is not populated through quantum fluctuations because $T_\mu^\mu=0$. 

If the system is strongly coupled there will also be an analogue of the notion of cross-section relevant for thermalisation which by large-$N$ counting scales as $\sigma \sim 1/(NT)^2$. 
The estimates for glueball DM in section \ref{sec:glueballs} are thus also valid for strongly coupled CFTs on the appropriate identification of the parameters.

We will focus in what follows on holographic realizations of CFTs. 

\subsection{Randall-Sundrum theories}

As is well known, a geometric realization of strongly coupled CFTs can be obtained through the AdS/CFT correspondence. 
At large $N$, the CFT has a dual description in terms of a weakly-coupled 5D gravity theory in AdS space of radius $L$. Coupling to 4D gravity is obtained 
introducing a UV brane that explicitly breaks conformal invariance. This is just the Randall-Sundrum 2 scenario \cite{Randall:1999vf}, see  \cite{ArkaniHamed:2000ds,Hebecker_2001} for the holographic interpretation. In this setup the coupling to gravity is uniquely determined. As explained above this corresponds to the fact that the minimal theory 
does not contain dimension 2 operators. Through the AdS/CFT dictionary $\Delta=2 \pm \sqrt{4+ M^2 L^2}$ a dimension 2 operator would correspond to 
a bulk scalar of mass $M^2=-4/L^2$ at the Freedman-Breitenloner unitarity bound. 

From the RS point of view the production of CFT states corresponds to the emission of 5D gravitons into the fifth dimension. 
The inclusive cross-section can be conveniently computed from the central charge of the dual CFT \cite{Penedones:2016voo}:
\begin{equation}
c_{\rm RS}= 320 \pi^2 M_5^3 L^3\,, ~~~~~~~~~~~~~~~~\Mpl^2=\frac{M_5^3 L}2+M_0^2\,,
\end{equation}
where $M_0^2$ is the UV contribution to the graviton kinetic term.
The 5D picture also allows us to discuss thermalisation effects that correspond to gravitational interactions in 5D. 
The effective coupling $\alpha_{\rm RS}\propto 1/c_{\rm RS}$ is suppressed in the regime where gravity is weakly coupled.  

Since we are ultimately interested in a theory with a mass gap we also introduce an IR brane so that the spectrum of excitations is discrete.
This can be realized with the Golberger-Wise stabilization mechanism \cite{Goldberger:1999uk}. This corresponds to the addition of a close-to-marginal deformation
to the CFT,  $[{\cal O}]=4+\epsilon$, dual to a scalar in 5D with mass $m\ll 1/L$.  The potential energy of the scalar naturally drives the IR to a large separation from the UV, generating 
a hierarchy of scales very similar to the Coleman-Weinberg mechanism of radiative symmetry breaking \cite{Rattazzi:2000hs}. 
The lightest state is a scalar, the radion,  that corresponds in the 4D picture to the dilaton, the Nambu-Goldstone boson of spontaneously broken conformal invariance,
see \cite{Medina:2011qc,Brax:2019koq,McDonald:2010iq,McDonald:2010fe,Gherghetta:2010cq} for alternative realizations of DM in holographic models.

\paragraph{Confined phase:}~\\

The low energy effective action is essentially fixed by the symmetries and can be cast in the following form \cite{Agashe:2019lhy}
\be\label{lagrangian-normalized}
\mathscr{L}= \frac {N^2}{16\pi^2}[ (\partial \varphi)^2 -  \hat V(\varphi)]+V_0\,,\quad
\hat V(\varphi)= \lambda_0 \varphi^4 \bigg[ 1 -\frac{4}{4+\epsilon} \bigg(\frac{\varphi}{f}\bigg)^\epsilon\bigg]+{\cal O}(\lambda_0^2)\,,
\ee
where $V_0=-\frac {N^2}{64\pi^2} \frac {\epsilon}{1+\epsilon/4} \lambda_0 f^4$ cancels the cosmological constant in the true vacuum. Above we introduce the effective number of degrees of freedom related to the central charge $c_{\rm RS}\sim N^2$.
Expanding around the minimum $\varphi=f$ and going to canonical normalization we find
\begin{equation}
\mathscr{L}= (\partial \chi)^2+   \lambda_0 \epsilon f^4 \left[2 \frac {\chi^2}{f^2} + \frac {40\pi}{N} \frac {\chi^3}{f^3}+ \frac {88\pi^2}{3 N^2} \frac {\chi^4}{f^4}+ \frac {64\pi^3}{5 N^3} \frac {\chi^5}{f^5}+\dots \right]\,.
\end{equation}

\begin{figure}[t]
\centering
\includegraphics[width=0.7\linewidth]{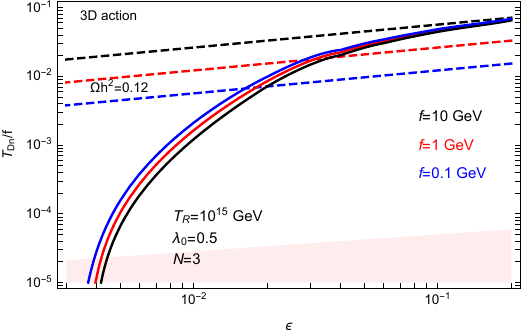}
\caption{\label{fig:TnCFT}\it Nucleation temperature in units of $f$ as a function of the anomalous dimension of the operator $[{\cal O}]=4+\epsilon$. Solid lines corresponds to tunnelling transitions computed with the 3d euclidean action, that we find dominant in the region of interest. The main difference from Ref. \cite{DelleRose:2019pgi} is due to the fact that the transition takes place during radiation domination with Hubble dominated by the SM degrees of freedom. DM abundance is reproduced where dashed lines intersect the nucleation curves. We neglect effects of cannibalism as they give small corrections. In the shaded region the dark sector vacuum energy dominates the energy budget.}
\end{figure}
The main difference from the glueball effective theory is that the potential is suppressed by the small parameter $\lambda_0 \epsilon$.
The interactions moreover are suppressed in the large $N$ limit, so that number-changing interactions are inefficient in the confined phase. If the system reaches thermal equilibrium in the deconfined phase we can compute the abundance as in eq. (\ref{DG-no-interactions}) with $L_h=V_0$ and neglecting the energy of the thermal bath since the phase transition occurs with some amount of supercooling. One finds
\be
\label{eq:dilatonabundance}
Y_{\rm \chi}=\frac{L_h/M}{s(T)}\bigg|_{T_n} = \sqrt{\epsilon \lambda_0}\frac {N^2}{64 \pi^2}\frac{45}{2\pi^2 g_*} \left(\frac {f}{T_{n}}\right)^3 =0.016 \sqrt{\epsilon \lambda_0}\frac {N^2}{64 \pi^2}\frac{45}{2\pi^2 g_*}\left(\frac {T_R}{M_p}\right)^{9/4} \left(\frac {f}{T_{Dn}}\right)^3\,,
\ee
so that
\begin{equation}
\frac{\Omega  h^2}{0.12}\approx \frac {\epsilon \lambda_0 f}{\rm GeV}\left(\frac  N 5\right)^2  \left(\frac{T_R} {10^{15}\,{\rm GeV}}\right)^{9/4}\left(\frac {f}{10 T_{Dn}}\right)^3\,.
\end{equation}

\paragraph{Phase transition:}~\\
The final abundance depends on the details of the phase transition through the amount of supercooling $T_n/f$. In this setup the phase transition is as follows \cite{Creminelli:2001th}: at high temperature the system is in the ``deconfined'' phase whose holographic 
description corresponds to an AdS black hole\footnote{The gravitational production of the CFT leads to a system out of thermal equilibrium. 
The process of thermalisation corresponds to the formation of a black hole in the 5D geometry.}. At low temperatures the configuration with IR brane has a lower free energy  and becomes favoured.
The two phases are separated by a first order phase transition.  The free energy of the decofined phase can be  estimated with $F\approx - N^2 T^4$, from which the critical temperature of the phase transition is $T_c^4/f^4\approx \epsilon \lambda_0/\sqrt{16\pi^2}$.  For the thermal CFT we consider a potential $V =N^2  \left( 4 \varphi^3 T + 3 \varphi^4 \right)$,  as suggested in \cite{Creminelli:2001th}.

Since the dark sector is in thermal equilibrium, the vacuum decay rate is $\Gamma \sim T_D^4 \exp(-S_E)$, where  $S_E$ stands either for the $O(4)$ symmetric bounce action or the $O(3)$ symmetric finite temperature Euclidean action. The nucleation temperature is obtained by solving $\Gamma=H^4$. In contrast to typical holographic phase transitions, in our model the Hubble rate is dominated
by the visible sector because the dark sector is always a small fraction of the total energy budget. In particular, the transition takes place during radiation domination when the Hubble parameter is dominated by the  degrees of freedom in the visible sector.
The nucleation temperature is thus determined by
\begin{equation}
S_E(T_D)=4 \log \frac {\Mpl}{T_D}-8 \log \frac T {T_D}+ \kappa \approx 4 \log \frac {\Mpl}{T_D}-6 \log \frac {\Mpl} {T_R}\,,
\end{equation}
where $\kappa $ is an order 1 number encoding various uncertainties, and in the second step we have used the relation $T\approx (\Mpl/T_R)^{3/4} T_D$.
This means that the transition requires a smaller Euclidean action, effectively delaying it compared to the standard scenario. Moreover, in our case, since the scale invariance is only broken slightly by the scaling dimension $\epsilon$, the bounce action (at low temperatures) has the approximate form
\footnote{At weak coupling this is
precisely the value of zero temperature and finite temperature Euclidean actions. At strong coupling 
the bounce action has a slightly different form that however does not change the qualitative behaviour discussed here. More details can be found in \cite{DelleRose:2019pgi}.}
\begin{equation}
S_E(T)=\frac {A}{\log M/T}\,.
\end{equation}
The slow logarithmic decrease of the action signals a transition with a potentially sizeable amount of supercooling, as expected in theories that are approximately conformal \cite{Creminelli:2001th,Randall:2006py,Konstandin:2011dr,Ellis:2018mja,Ellis:2019oqb}.
The dark nucleation temperature
is
\begin{equation}
T^n_D=\sqrt{M \Mpl} \left(\frac {T_R}{\Mpl}\right)^{\frac 3 4} e^{-\frac 1 4\sqrt{4 A +\log^2 \frac {M^2 \Mpl}{T_R^3}}}\,.
\end{equation}

If $A$ is negligible, the transition takes place when $T^n_D\approx M$, i.e. as soon as the mass scale is reached. Note that in this case the latent heat of the transition
is much smaller than the energy density. If $A$ is large, the temperature of the dark sector drops to very small values and the thermal energy becomes subdominant compared
to the vacuum energy.  

 
 If Fig. \ref{fig:TnCFT} we present an example of the nucleation temperature following \cite{DelleRose:2019pgi}.
In order to reproduce the abundance of DM the phase transition must be fast, requiring a significant deviation from conformality.

\paragraph{Small gravity wave signals:}~\\
The above computation allows us to extract the parameters relevant for the computation of the gravitational wave spectrum, that originates at the phase transition. Here we refer the reader to  \cite{Breitbach:2018ddu} for gravitational wave signatures from secluded dark sectors.
The parameters of the phase transition are simply extracted by noting that the temperature of the dark sector is linearly related to the SM temperature,
$T_D\approx (T_R/\Mpl)^{3/4} T$. In particular  $\beta/H\equiv T\frac{dS}{dT}|_{T_{D}^n}$ is not modified compared to the standard case with same temperature. 
In the case where a sizeable supercooling is present, the amplitude of the gravitational wave power spectrum is
\begin{equation}\label{gw-spectrum}
\Omega_{\rm GW} h^2 \approx 5 \times 10^{-6} \left(\frac {H_*}{\beta}\right)^2  \left(\frac {T_R}{\Mpl}\right)^3\,.
\end{equation}
This value gives an upper bound on the amplitude of the gravity waves produced: if the dark sector energy density is a subdominant component of the total energy the amplitude is further suppressed by the squared ratio of the latent heat and the SM energy density. Given the already large suppression in eq.~\eqref{gw-spectrum}, due to the different temperatures, which already produces an amplitude too small to be seen in current experiments, the gravity wave background is thus negligible.
If the system has not reached thermal equilibrium a quantitive study of the phase transition is more difficult. We believe that the physics is roughly the following:
as long as the density is larger than $\Lambda^3$ the system remains in the deconfined phase. 
When the density drops below this critical value nothing prevents the system from relaxing to the ground state and the phase transition 
completes. Given the small amount of energy density in the dark sector we do not however expect a significant gravity wave signal also in this case.

\subsection{Holographic confining gauge theories}

A different realization of the same idea is the holographic dual of a confining gauge theory as introduced in  \cite{Polchinski:2000uf}.
In particular, due to supersymmetry it is possible to control the dynamics and have an explicit realization of the CFT, ${\cal N}=4$ super-Yang-Mills in the simplest case. 
The main difference from the previous discussion is that the deformation is strongly relevant, growing rapidly in the IR. Because the explicit breaking of conformal invariance is large,
it is not possible to identify a dilaton-like particle. This model at finite temperature was studied in \cite{Freedman:2000xb}. 
As in the Randall-Sundrum model, the high-temperature regime is described by a 5D geometry with a black-hole or black brane. 
As the system evolves to lower temperatures, the confining phase becomes favoured. Due to the absence of a light dilaton it is not possible to compute the transition rate 
within effective field theory control. Nevertheless, since the deformation is strongly relevant, we expect the rate to be fast as in QCD-like theories so that $T_n\sim T_c$.

\section{Conclusions}\label{sec:conclusions}

The possibility that DM interacts with the SM only through gravitational interactions is both compelling and concerning.
A truly dark sector naturally contains cosmologically stable DM candidates and would automatically escape any laboratory experimental bound unless the mass were exceedingly small.
While the observational prospects of DM look very dim in this case, one might nevertheless seek to identify the most plausible scenarios.
In light of that, it is interesting to consider the minimal production mechanism due to the coupling to gravity that unavoidably populates the dark sector. 
One firm prediction is that the abundance of DM requires a large reheating temperature that in turn implies an inflationary scale not very far from the current bound.
Therefore  discovering a large scale of inflation through gravity waves would lend credence to this scenario.

In this work we investigated the population of the dark sector in the relativistic regime where the results become model independent:
with the notable exception of  elementary scalars and gravitons, due to approximate conformal invariance, production through quantum fluctuations is suppressed
and the abundance of the dark sector particles is determined by the central charge of the dark sector, roughly a measure of the number of degrees of freedom.
Gravitational production leads to a dark sector plasma strongly out of equilibrium with small numerical density compared to the typical energy. 
Depending on interactions very different dynamics take place that modify the nature of DM and its abundance.
In particular when the interactions are strong and the mass is light the system thermalises at a temperature much lower than the SM and begins its own thermal history.

Gravitational production appears particularly attractive for the DM glueball scenario. Such a situation requires the temperature of the dark sector to be much 
smaller than that of the SM to escape experimental constraints. This requirement is automatically satisfied with gravitational production because the dark sector is underabundant.
The dynamics lead to interesting effects such as out-of-equilibrium phase transitions and cannibalism.  
Similar effects can be studied in strongly coupled CFTs and their holographic dual theories where for example the DM is a light dilaton from spontaneous breaking of 
conformal invariance.

There are several questions that we are planning to pursue in future work. As we have argued, production through quantum fluctuations is suppressed in our scenarios.
It will nevertheless be productive to investigate the contribution to the abundance from small deviations from conformality. Gravitational production is also relevant 
for other sectors such as baryon-like DM. Finally it is interesting to note that gravitons are also produced with a yield and distribution determined in this paper.

{\small
\subsubsection*{Acknowledgements}
This work is supported by MIUR grants PRIN 2017FMJFMW and 2017L5W2PT and INFN grant STRONG. HT thanks the Science and Technology Facilities Council (STFC) for a postgraduate studentship.
We would like to thank Andrea Cappelli, Raghuveer Garani for discussions, and the Galileo Galilei Institute for hospitality during this work.}

\appendix

\section{CFT production}
\label{app:A}

In this appendix we derive  the relevant formulas for the thermal production of CFT states.
All the results depend solely on the central charge of the CFT.

In general a CFT is endowed with a traceless energy momentum tensor $T_{\mu\nu}$ of dimension 4 so that the minimal coupling to gravity is just $g_{\mu\nu}T^{\mu\nu}$. 
In strongly interacting CFTs, modulo the highly plausible assumption $T_\mu^\mu=0$ \cite{Luty:2012ww}, the leading coupling to gravity is uniquely determined, $h_{\mu\nu}T^{\mu\nu}$.\footnote{Tracelessness of $T_{\mu\nu}$ can be violated
only if dimension 2 operators exist allowing for the marginal coupling $R \cal{O}$. As explained below this is realized in theories with weakly coupled scalars.} 
The 2-point function of $T_{\mu\nu}$ is in turn fixed by conformal invariance as
\begin{equation}
\langle T_{\mu\nu}(x) T_{\rho \sigma }(0)\rangle = \frac 1{4\pi^4} P_{\mu\nu  \sigma\rho} \frac c {x^8}\,, \quad P_{\mu\nu\sigma\rho}=\frac 1 2(I_{\mu\sigma} I_{\nu\rho}+I_{\mu\rho}I_{\nu\sigma})-\frac 1 4 \eta_{\mu\nu}\eta_{\sigma\rho} \,,\quad I_{\mu\nu}= \eta_{\mu\nu}- 2 \frac {x_\mu x_\nu}{x^2}\,,
\end{equation}
where $c$ is the central charge of the CFT, which roughly measures the number of degrees of freedom of the theory.
In momentum space the 2-point function has a dependence $p^4 \log p$. More precisely with our normalizations \cite{Gubser:1997se},
\begin{equation}
\langle T_{\mu\nu}(p) T_{\rho \sigma }(-p)\rangle= \frac{c}{7680 \pi^2}  \left(2\pi_{\mu\nu} \pi_{\rho\sigma}-3\pi_{\mu\rho} \pi_{\nu\sigma}-3\pi_{\mu\sigma} \pi_{\nu\rho}\right)  \log(-p^2)\,,
\end{equation}
where $\pi_{\mu\nu}=\eta_{\mu\nu}p^2-p_\mu p_\nu $. The tree level graviton propagator can be chosen as
\begin{equation}
\langle h_{\mu\nu}(p) h_{\rho \sigma }(-p)\rangle = P_{\mu\nu\rho\sigma} \frac {i}{p^2+i \epsilon}\,,~~~~~~~~~~~~P_{\mu\nu\rho\sigma}= \frac 1 2(\eta_{\mu\rho}\eta_{\nu \sigma}+\eta_{\mu\sigma}\eta_{\nu \rho}-\eta_{\mu\nu}\eta_{\rho \sigma})\,.
\end{equation}
The 1-loop correction to the graviton propagator is thus
\begin{equation}
P^{\mu\nu\alpha\beta} \langle T_{\alpha\beta}(p) T_{\gamma \delta}(-p)\rangle P^{\gamma\delta\rho\sigma} =\frac 1 {\Mpl^2} \frac{c}{7680\pi^2}\left(2\pi_{\mu\nu} \pi_{\rho\sigma}-3\pi_{\mu\rho} \pi_{\nu\sigma}-3\pi_{\mu\sigma} \pi_{\mu\rho}\right)  \log(-p^2)\,.
\end{equation}
It is now simple to extract the inclusive cross-section for tree-level production of CFT states.  To achieve this
we can use the optical theorem so that the inclusive cross-section  is proportional to the imaginary part of the graviton propagator, i.e. 
to the central charge $c_f$ appearing in the 2-point function of the energy momentum tensor of the CFT. The opposite process of production of SM states is instead proportional to $c_i$.
One finds that the annihilation cross-section, averaged over initial states, is given by
\footnote{We quote the cosmological cross-section entering Boltzmann equations that differs by a factor 2 for Dirac particles compared to \cite{Tang:2017hvq}.}
\begin{equation}
\sigma_{\rm tot} =  \frac{c_i c_f}{g_i^2}\frac {1}{10240\pi}  \frac{s}{M_p^4}\,,
\label{eq:sigmaCFT}
\end{equation}

For production then the only difference between weak coupling and strong coupling is the value of the central charge. 
For graviton production using the result in \cite{Donoghue:1994dn} we find that the same formula applies with $c_D=28$,
if produced from the conformally coupled states.

This zero temperature cross-section is the building block to determine the gravitational production of the CFT.

\paragraph{Boltzmann equations:}~\\
To compute the distribution function of the dark sector we follow \cite{Dvorkin:2019zdi,Bae:2017dpt}.
In general the Boltzmann equation for the phase space distribution of particles in the plasma reads
\begin{equation}
\frac {\partial f}{\partial t}- H \frac {|\vec p|^2}{E} \frac {\partial f}{\partial E}= \frac {C[f]}E\,,
\label{eq:Boltzmann}
\end{equation}
where $C[f]$ is the collision term. For $2\to 2$ processes this is given by
\begin{equation}
\begin{split}
\frac {C(t,p)}{E} &= \frac 1 {2E} \int \Pi_{i=1}^3 \frac {d^3 p_i}{(2\pi)^3 2 E_i} (2\pi)^4 \delta^4(p_1+p_2+p_3+ p) \sum |{\cal M}|^2\\
&\times \left[f_1(p_1) f_2(p_2)(1\pm f_3(p_3))(1\pm f(p))-f_3(p_3)f(p)(1\pm f(p_1))(1\pm f_2(p_2))\right]\,.
\end{split}
\end{equation}

We are interested in the thermal production of an underpopulated  sector so that $f_{1,2}=f_{eq}$ and $f_{3,4}\approx 0$. The exact solution of (\ref{eq:Boltzmann})
is given by
\begin{equation}
f(t, p)= \int_{t_i}^t dt \frac 1 E C\left(t, \frac {a(t)}{a(t')} p\right)\,.
\end{equation}

The collision term can be written as
\begin{equation}
\frac {C(t,p)}{E} =\frac 1 {2E} \int \Pi_{i=1}^3 \frac {d^3 p_i}{(2\pi)^3 2 E_i} (2\pi)^4 \delta^4(p_1+p_2+p_3+ p)\sum |{\cal M}|^2\times \left[f_{eq}(p_1) f_{eq}(p_2)\right]\,.
\end{equation}

Using the identity
\begin{equation}
f_{eq}(p_1) f_{eq}(p_2)=(1+f_{eq}(p_1))(1+ f_{eq}(p_2))e^{-\frac {E_3+E_4}T}
\end{equation}
in the massless limit and neglecting the quantum statistics we obtain
\begin{equation}
\frac {C(t,p)}{E}\approx \frac  {e^{-p/T}}{512 \pi^3 p^2} \int ds \int_{\frac {s} {4 p}}^{\infty} dp_3 \frac 2 s e^{-p_3/T}  16\pi s\,\sigma(s)=
T \frac  {e^{-p/T}}{16 \pi^2 p^2} \int ds\, s e^{-s/(4 p T)} \sigma(s)\,.
\end{equation}
Inserting (\ref{eq:sigmaCFT}) we obtain
\begin{equation}
\frac {C_{\rm SM+SM \to CFT}}{E}=\frac{  c_i c_f }{1280 \pi^3}  \frac {T^4}{\Mpl^4} p e^{-p/T}\,.
\label{eq:collision}
\end{equation}
From this we can compute the thermally averaged cross-section as
\begin{equation}
\langle \sigma v\rangle= \frac 1 {n_{eq}^2}\int \frac{d^3 p}{(2\pi)^3} \frac {C_{\rm SM+SM \to CFT}}{E}= \frac {c_i c_f}{g_i^2} \frac {3}{1280\pi} \frac {T^2}{\Mpl^4}\,,
\end{equation}
where $n_{eq}= g_i T^3/\pi^2$. This result agrees with the standard formula \cite{Gondolo:1990dk}. The only approximation made to
compute the thermally averaged cross-section is the neglecting of the quantum corrections which leads to ${\cal O}(10\%)$ error in the relativistic regime. 
From eq. (\ref{eq:collision}) we derive the energy distributions as
\begin{equation}
f(T,p)=\int_{T}^{T_R} \frac {dT'}{T' H(T')} \frac {C(T',\frac{p\,T'} T)}{p}=\frac{c_D c_{\rm SM}}{\sqrt{g_*}}\frac {1} {128\sqrt{10}\pi^4} \left(1-\frac {T^3}{T_R^3}\right)\frac {T_R^3}{\Mpl^3} \frac p T  e^{-p/T}\,.
\end{equation}
From this we can finally compute the number and energy densities
\begin{equation}
n(T)= \frac {3} {128\sqrt{10}\pi^6} \frac{c_D c_{\rm SM}}{\sqrt{g_*}} \left(1-\frac {T^3}{T_R^3}\right)\frac {T_R^3}{\Mpl^3} T^3\,,\quad\quad \rho(T)= 4 T n(T)\,.
\end{equation}
in agreement with eq. (\ref{eq:yield}).

\pagestyle{plain}
\bibliographystyle{jhep}
\small
\bibliography{biblio}

\end{document}